\documentclass[twocolumn,prc,showpacs,amsmath]{revtex4}

\usepackage{graphicx}
\usepackage{dcolumn}
\usepackage{bm}

\begin{document}


\title{DEUTERON TENSOR POLARIZATION COMPONENT ${T_{20}(Q^2)}$
AS A CRUCIAL TEST FOR DEUTERON WAVE
FUNCTIONS}

\author{A.F.~Krutov}

\affiliation{Samara State University, 443011 Samara, Russia}
\email{krutov@ssu.samara.ru}

\author{V.E.~Troitsky}
\affiliation{D.V.~Skobeltsin Institute of Nuclear Physics,
  Moscow State University, Moscow 119992, Russia}
\email{troitsky@theory.sinp.msu.ru}
\date{\today}

\begin{abstract}
The deuteron tensor polarization component ${T_{20}(Q^2)}$ is calculated
by relativistic Hamiltonian dynamics approach. It is shown that in
the range  of momentum transfers available in to-day experiments,
relativistic effects, meson exchange currents and the choice of
nucleon electromagnetic form factors almost do not influence the value of
${T_{20}(Q^2)}$. At the same time, this value depends strongly on the
actual form of the deuteron wave function, that is on the model of
$NN$--interaction in deuteron. So the existing data for
${T_{20}(Q^2)}$ provide a crucial test for deuteron wave
functions.
\end{abstract}


\maketitle

\section{Introduction}
\label{sec:intro}

It is for a long time that in the framework of nonrelativistic approaches
deuteron tensor polarization was being considered
as an important tool to probe  nucleon--nucleon interaction at
short distances
(see, e.g., Refs.~\cite{Lev73,MoG74}), giving a possibility to choose
between different model deuteron wave functions, that is between different
models of
$NN$--interaction. During last years great success was achieved in
polarization experiments on the elastic electron-deuteron
scattering~\cite{ScB84, Dmi85, GiH90, ThA91, FeB96,
BoA99,AbA00}. Now accessible values of $Q^2$ are so large that the
 relativistic theory is needed. Unfortunately, further progress in
 these measurements is questionable  because no reasonable technique
exists to extend polarization measurements to higher $Q^2$ ~\cite{Gil04,
GiG02}. So, in the immediate future one does not ezpect any new
experimental information on the subject in question.

In the present work we analyze
the existing data on polarization
$ed$-scattering in connection with different $NN$--interaction models on
the base of an essentially relativistic approach. We show that the existing
data provide the crucial test of the deuteron wave function even in the
relativistic theory. This possibility is based on the following results
obtained in the paper.
\begin{itemize}
\item  The relativistic corrections to $T_{20}(Q^2)$  are small up to $Q^2
\simeq 3$(GeV/c)$^2$.
\item  The quantity $T_{20}(Q^2)$  is almost independent of the actual
nucleon electromagnetic form factors (although the deuteron structure
functions $A(Q^2)$  and $B(Q^2)$ depend strongly).
\item  The contribution of meson exchange currents (MEC) to
$T_{20}(Q^2)$ are small.
\item  The quantity $T_{20}(Q^2)$ depends strongly on the choice of the
deuteron wave function, that is on the model of
$NN$--interaction.
\end{itemize}
We consider the most popular model deuteron wave functions to obtain the
best description of polarization data.

 The analysis is performed
in the framework of the  variant of the
instant form of the relativistic Hamiltonian dynamics (IF RHD)
developed by the authors~\cite{BaK95, KrT02,
KrT03prc, KrT03}.(RHD is sometimes called Poincar{\'e} invariant
quantum mechanics (see, e.g.,~\cite{KeP06}.)) The main features of our
approach to deuteron are the following. First, the form of the dynamics is
close to nonrelativistic case. Second, our method of construction of the
matrix element of the electroweak current operator makes it possible to
formulate relativistic impulse approximation in such a way that the
Lorentz covariance of the current is ensured. In our approach it is
possible to use the Siegert theorem~\cite{Sie37,Bai79} to estimate the
contribution of meson exchange currents to the deuteron electromagnetic
structure. Our estimation of the role of different contributions --
nucleon dynamics, relativistic effects, meson exchange currents, nucleon
internal structure -- demonstrates that one can use the function
$T_{20}(Q^2)$ to discriminate different model deuteron wave functions and
to choose the most adequate models of nucleon--nucleon interaction. Our
calculation shows that the most popular model wave functions~\cite{LaL81,
StK94, Mac01} do not give adequate description of $T_{20}(Q^2)$ and are to
be discriminate in favor of those obtained in the dispersion potentialless
inverse scattering approach with no adjustable parameters~\cite{MuT81,
Tro94} and giving the best description.

The paper is organized as follows. In Section 2 we formulate the problem
of obtaining the best deuteron wave function using the data on
$T_{20}(Q^2)$. Section 3 contains a brief review of IF RHD.
In Section 4 the electromagnetic deuteron form factors are calculated and
the deuteron tensor polarization
${T_{20}(Q^2)}$ is given in terms of these form factors. The relativistic
effects and the effect of MEC are estimated. Section 6 presents the
conclusions. In the Appendix A(B) the equations for relativistic
(nonrelativistic) free form factors for two nucleons in $^3S_1- ^3D_1$
channel are given. These form factors enter the equations
(\ref{GqGRIP})  and (\ref{GqGNIP}) of the main text.

\section{
 Wave functions from
deuteron experiments}

The problem of obtaining the most adequate wave function from deuteron
experimental data, in general, can be correctly formulated only in the
conventional nonrelativistic nuclear model.
In the framework of this model all existing nucleon--nucleon interaction
potentials have the correctly fixed
long--distance part defined by the one--pion exchange and
the intermediate and the short--distance part of strongly pronounced model
character, different in different models. That is why the deuteron wave
functions in coordinate representation for different approaches usually
coincide at $r\;\geq\; 1.5$ fm  and differ essentially at $r\;\leq\;0.5$
fm. This quite obvious formulation of the problem in conventional
nonrelativistic nuclear model, however, is not valid out of the framework
of the approach. This takes place, for example, when relativistic effects
are taken into account, or meson exchange currents (interaction currents),
or different effects in deuteron electromagnetic structure caused by quark
degrees of freedom. These effects are usually strongly model dependent and
contain a kind of arbitrariness, so that they ``mask'' effectively the
dependence of observables on the choice of the dynamics of
$NN$--interaction. For example, the experimental data on $T_{20}(Q^2)$ in
$ed$-scattering were described well by the relativistic
approach of Ref.~\cite{VaD95} because of functional arbitrariness in the
definition of nucleon electromagnetic current. Another example is
represented by the calculation of the contribution of meson exchange
currents to deuteron electromagnetic from factors where in fact an
arbitrariness is contained in the $\rho\pi\gamma$- form
factor~\cite{GaH76}.

Now one understands clearly that it is impossible to neglect relativistic
effects in deuteron so that one needs a consistent
relativistic formulation of the deuteron problem, particularly at large
momentum transfer~\cite{GiG02, ChC88}.
As was noted in~\cite{GiG02} to-day there are two main classes of
relativistic schemas of the description of deuteron. The first class is
based on field--theoretical concepts ( following the paper~\cite{GiG02} --
propagator dynamics). This class contains the Bethe- Salpeter equation and
quasipotential approaches (and the approach ~\cite{VaD95}, too). The
second class -- relativistic Hamiltonian dynamics (RHD) -- is based on the
realization of the Poincar\'e algebra on the set of dynamical observables
of the system with the finite numbers of degrees of freedom. One can find
the description of RHD method in the reviews~\cite{KeP91} (see
also~\cite{Kli98}) and especially the case of deuteron in the
reviews~\cite{GiG02, GaV01}. As is noted in~\cite{GiG02}, the connection
between the propagator dynamics and RHD is ambiguous. Each of the
approaches has its own advantages as well as difficulties. One should
mention in addition the dispersion methods of
describing composite systems, these methods dealing with, in fact, finite
numbers of degrees of freedom, as RHD does~\cite{ TrS69, KiT75, AnK92,
AnM95}.

In the mentioned relativistic approaches the process of construction of
the operator of Lorentz covariant conserved electromagnetic current is
connected to the relativistic nucleon-nucleon dynamics used in the
approach. That is why the problem of obtaining the information about the
dynamics itself from the deuteron data is not, in general, correctly
formulated.

Is it possible in principle to formulate correctly the problem of
obtaining the most adequate deuteron wave functions from the deuteron
experiments? Our opinion is that it is possible if the following
 requirements are satisfied.

1. It is necessary to find an approach
relativistic from the very beginning with the dynamics close to
nonrelativistic Schr\"odinger dynamics, that is with relativistic deuteron
wave functions which are close to the nonrelativistic ones.

2. It is necessary to find such a measurable quantity that, in the chosen
approach, is almost independent of relativistic corrections, meson
exchange currents and the internal structure of nucleons.

In this paper we propose  such a relativistic approach -- a variant of IF
RHD developed by the authors in Refs.~\cite{BaK95,KrT02, KrT03prc,
KrT03}. In this approach the adequate observable is the
component $T_{20}(Q^2)$ of the deuteron polarization tensor in elastic
electron--deuteron scattering.

Let us review briefly the dynamics in our relativistic
approach.

\section{Dynamics in IF RHD}

We use the so called instant form of relativistic Hamiltonian
dynamics (IF RHD) ~\cite{Dir49}. In this form the kinematic subgroup of
Poincar\'e algebra contains the generators of the group of rotations and
translations in the three--dimensional Euclidean space (interaction
independent generators):
\begin{equation}
 \hat{\vec J}\>,\quad\hat{\vec P}\;.
\label{kinem}
\end{equation}
The remaining generators of the time translation and Lorentz
boosts are Hamiltonians (interaction depending):
\begin{equation}
\hat P^0\>,\quad \hat{\vec N}\;. \label{hamil}
\end{equation}

The additive including of interaction into the mass square
operator (Bakamjian--Thomas procedure, see, e.g.,~\cite{KeP91}
for details) presents one of the possible technical ways to
include interaction in the algebra of the Poincar\'e group:
\begin{equation}
\hat M_0^2 \to \hat M_I^2 = \hat M_0^2 + \hat U \;. \label{M0toMI}
\end{equation}
Here $\hat M_0$ is the operator of invariant mass for the free
system and $\hat M_I$ -- for the system with interaction. The
interaction operator $\hat U$ must satisfy the following
commutation relations:
\begin{equation}
\left [\hat {\vec P},\,\hat U\right ] = \left[\hat {\vec
J},\,\hat U\right ] = \left [\vec\bigtriangledown_P,\,\hat
U\right ] = 0\;. \label{[PU]=0}
\end{equation}
These constraints (\ref{[PU]=0}) ensure that the algebraic
relations of Poincar\'e group are fulfilled for the interacting
system. The relations (\ref{[PU]=0}) mean that the interaction
potential does not depend on the total momentum of the system as
well as on the projection of the total angular momentum. The
conditions (\ref{M0toMI}) and (\ref{[PU]=0}) can be considered only as
the model ones. There are other approaches  with
potential depending on the total momentum but they are out
of scope of this paper.

In RHD the wave function of the system of interacting particles
is the eigenfunction of a complete set of commuting operators. In
IF this set is:
\begin{equation}
 {\hat M}_I^2\>,\quad
{\hat J}^2\>,\quad \hat J_3\>,\quad \hat {\vec P}\;.
\label{complete}
\end{equation}
${\hat J}^2$ is the operator of the square of the total angular
momentum. In IF the operators ${\hat J}^2\;,\;\hat J_3\;, \;\hat
{\vec P}$ coincide with those for the free system. So, in
Eq.(\ref{complete}) only the operator $\hat M_I^2$ depends on the
interaction.

To find the eigenfunctions of the system (\ref{complete}) one
has first to construct the adequate basis in the state space of
composite system. In the case of two-particle system (for
example, two-nucleon system) the Hilbert space in RHD is the
direct product of two one-particle Hilbert spaces: ${\cal
H}_{NN}\equiv {\cal H}_N\otimes {\cal H}_{N}$.

As a basis in ${\cal H}_{NN}$ one can choose the following set of
two-particle state vectors where the motion of the two-particle
center of mass is separated and where three operators of the set
(\ref{complete}) are diagonal:
\begin{equation}
|\,\vec P,\;\sqrt{s},\;J,\;l,\;S,\;m_J\,\rangle\;, \label{PkJlSm}
\end{equation}
here $P_\mu = (p_1 +p_2)_\mu$, $p_1^2 = p_2^2 =M^2$, $M$ is
nucleon mass, $P^2_\mu = s$, $\sqrt {s}$ is the invariant mass of
the two-particle system, $l$ -- the orbital angular momentum in
the center-of-mass frame (C.M.S.), $\vec S\,^2=(\vec S_1 + \vec
S_2)^2 = S(S+1)\;,\;S$ -- the total spin in C.M.S., $J$ -- the
total angular momentum with the projection $m_J$, the parameters
$S$ and $l$ play the role of invariant parameters of degeneracy.

As in the basis (\ref{PkJlSm}) the operators ${\hat J}^2\;,\;\hat
J_3\;,\;\hat {\vec P}$ in (\ref{complete}) are diagonal, one
needs to diagonalize only the operator $\hat M_I^2$   in order to obtain
the system wave functions.

The eigenvalue problem for the operator ${\hat M}^2_I$ in the
basis (\ref{PkJlSm}) coincides with the nonrelativistic
Schr\"odinger equation within following difference between
corresponding eigenvalues (see,  e.g.,~\cite{ChC88, KeP91}):
\begin{equation}
\left(\frac{M^2_d}{4\,M} -M\right) - (M_d - 2M) = \frac{(M_d -
2M)^2}{4\,M} =  \epsilon^2_d/4M\;. \label{epsilond}
\end{equation}
Here $M_d$ is the deuteron mass,  $\epsilon_d$ is the deuteron binding
energy.

The difference (\ref{epsilond}) is negligible for most problems.

The corresponding composite--particle wave function has the form
\begin{eqnarray}
\langle\vec P\,',\,\sqrt {s'},\,J',\,l',\,S',\,m_J'|\,p_c\rangle
&&\nonumber \\
=N_C\,\delta (\vec P\,' - \vec p_c)\delta _{JJ'}\delta
_{m_Jm_J'}\, \varphi^{J'}_{l'S'}(k')\;,&& \label{wf}
\end{eqnarray}
$|\,p_c\rangle$ is an eigenvector of the set (\ref{complete});
$J(J+1)$ and $m_J$ are the eigenvalues of $\hat J^2\,,\;\hat
J_3\,,$ respectively. $N_C$ is the normalization constant.

We use the normalization with the relativistic density of states:
\begin{equation}
k^2\,dk\quad\to\quad \frac{k^2\,dk}{2\sqrt{(k^2 + M^2)}}\;.
\label{rel den}
\end{equation}

This gives the
following two--particle wave function of relative motion for equal
masses and total angular momentum and total spin fixed:
\begin{equation}
\varphi^J_{lS}(k(s)) = \sqrt[4]{s}\,u_l(k)\,k\;, \label{phi(s)}
\end{equation}
with the normalization condition:
\begin{equation}
\sum_l\int\,u_l^2(k)\,k^2\,dk = 1\;. \label{norm}
\end{equation}
Functions $u_l(k)\;,\;l=0\;,\;2$ coincide with the model
nonrelativistic deuteron wave functions within the difference
(\ref{epsilond}). The wave function (\ref{phi(s)}) coincides
with that obtained by ``minimal relativization'' in~\cite{FrG89}.

So, in our approach the wave functions in the RHD sense are close to the
corresponding nonrelativistic wave functions and the dynamical equation is
close to the nonrelativistic Schr\"odinger equation.

Let us emphasize that our formalism enables one to use any model
wave functions obtained as the solution of
Schr\"odinger equation.

In this paper we consider the following models of $NN$--interaction:
Paris potential~\cite{LaL81}, the versions
I, II and 93 of the Nijmegen model~\cite{StK94},
charge--dependent version of Bonn potential~\cite{Mac01}. The
deuteron wave functions for these potentials give the results for
deuteron electromagnetic properties that differ essentially from
one another. It is a difficult task to give the preference to one
of them. Quite different kind of results presents the deuteron wave
functions (MT)~\cite{MuT81} obtained in potentialless
approach to the inverse scattering problem (see for the details~\cite{Tro94}).

Now let us calculate the deuteron electromagnetic
form factors.

\section{Deuteron electromagnetic form factors}

The main point of our approach is a kind of construction of
the matrix element of electroweak current operator.  In our method the
electroweak current matrix element satisfies the relativistic
covariance conditions and in the case of electromagnetic
current also the conservation law automatically.  The properties
of the system as well as the approximations are formulated in
terms of form factors.  The approach makes it possible to
formulate relativistic impulse approximation in such  a way that
Lorentz covariance of the current is ensured.  In the
electromagnetic case the current conservation law is also ensured.

Usually it is supposed that it is necessary to take MEC into
account in order to provide the gauge invariance and the current
conservation~\cite{KeP91}. However to--day the construction of the
relativistic impulse approximation without breaking of the
relativistic covariance and current conservation law is a common
trend of different approaches~\cite{GiG02,KrT02, Kli98, LeP00,
AlK01}. In our approach this is realized by making use of the
Wigner--Eckart theorem for the Poincar\'e group.  It enables one
(for given current matrix element) to separate the reduced matrix
elements (form factors) which are invariant under the Poincar\'e
group action.  The matrix element of a given operator is
represented as a sum of terms, each one of them being a covariant
part multiplied by an invariant part.  In such a representation
the covariant part describes the transformation properties of the
matrix element. The conservation law  is satisfied explicitly due
to the fact that the vector of the covariant part is orthogonal
to the vector $Q_{\mu}$. All the dynamical information on the
transition is contained in the invariant part (form factors). In
our variant of the impulse approximation (modified impulse
approximation) the reduced matrix elements are calculated with no
change of covariant part (see~\cite{KrT02} for the details) although
neglecting MEC. The correct transformation properties are thus
guaranteed.

Charge, quadrupole and magnetic form factors of deuteron in our approach
have the form~\cite{KrT03prc}:
$$
G_C(Q^2) = \sum_{l,l'}\int\,d\sqrt{s}\,d\sqrt{s'}\, \varphi^l(s)\,
g^{ll'}_{0C}(s\,,Q^2\,,s')\, \varphi^{l'}(s')\;,
$$
\begin{equation}
G_Q(Q^2) \!=\!
\frac{2\,M_d^2}{Q^2}\sum_{l,l'}\int\!\!d\sqrt{s}\,d\sqrt{s'}\,
\varphi^l\!(s)g^{ll'}_{0Q}(s,Q^2\!,s')\varphi^{l'}\!(s'),
\label{GqGRIP}
\end{equation}
$$
G_M(Q^2) \!=\!-M_D\!\sum_{l,l'}\int\!\!d\sqrt{s}\,d\sqrt{s'}\,
\varphi^l\!(s)g^{ll'}_{0M}(s,Q^2\!,s') \varphi^{l'}\!(s').
$$
Here $g^{ll'}_{0i}((s\,,Q^2\,,s')\;,\;i=C,Q,M$ are free
charge, quadrupole and magnetic two--particle form factors, that is the
form factors describing electromagnetic properties of the system of proton
and neutron without interaction, the system having deuteron quantum
numbers, $l\;,\;l'=$ 0,2, -- orbital moments, $\varphi^l(s)$ --
wave functions in the sense of RHD.

Free two--particle form factors for a system of two fermions with total
momentum 1 (without taking into account of $D$-state) were
obtained in~\cite{KrT03prc}.
Corresponding equations for the neutron--proton system with deuteron
quantum numbers are given in Appendix A. Free two--particle charge (only)
form factors of proton--neutron system without interaction in deuteron
quantum numbers channel are given also in~\cite{AfA98}.

For the deuteron electromagnetic form factors
(\ref{GqGRIP}) the correspondence principle is valid. The
nonrelativistic limit
($M\;\to\;\infty$) of Eq.(\ref{GqGRIP}) gives the standard equations
for deuteron form factors in nonrelativistic impulse approximation in
terms of wave functions in momentum representation
(see, e.g.,~\cite{BrJ76, MaZ78}):
$$
G^{N\!R}_C(Q^2)\!=\!\!\sum_{l,l'}\!\int\!k^2 dk\,k'\,^2 dk' u^l(k)
\tilde g^{ll'}_{0C}(k,Q^2\!,k') u^{l'}(k'),
$$
\begin{equation}
G^{N\!R}_Q(Q^2)\!=\!
\frac{2\,M_D^2}{Q^2}\!\sum_{l,l'}\!\!\int\!\!k^2 dk\,k'^2dk'
u^l(\!k)\tilde g^{ll'}_{0Q}\!(k\!,Q^2\!\!,k') u^{l'}\!\!(k'),
\label{GqGNIP}
\end{equation}
$$
G^{N\!R}_M(Q^2)\!=\!-M_D\!\!\sum_{l,l'}\!\!\int\!\!k^2 dk\,k'\,^2dk'
u^l(\!k)\tilde g^{ll'}_{0M}\!(k\!,Q^2\!\!,k') u^{l'}\!\!(k').
$$
Free charge, quadrupole and magnetic two--particle form factors $\tilde
g^{ll'}_{0i}(k\,,Q^2\,,k')\;,\;i=C,Q,M$ can be calculated as
nonrelativistic limits of relativistic two--particle form factors given in
Appendix A. The explicit forms of free nonrelativistic two--particle form
factors are given in Appendix B.

So, to solve the problem in question we propose the essentially
relativistic approach which gives a possibility of calculation of deuteron
electromagnetic form factors and takes into account the relativistic
covariance and the conservation law for the electromagnetic current. The
efficiency of our approach was demonstrated in a number of
calculations~\cite{BaK95, KrT02,KrT03prc,KrT03}. In particular, the values
of neutron charge form factor extracted from the deuteron charge form
factor \cite{KrT03} are in good accordance with the values of other
authors.

Now let us apply our formalism to polarization $ed$-scattering.

\section{Polarized $ed$-scattering}

The component $T_{20}(Q^2)$  of the deuteron polarization tensor in
elastic $ed$-scattering can be written in terms of deuteron form factors
(\ref{GqGRIP})
in the following form~\cite{GiG02}:
\begin{equation}
T_{20}(Q^2) = -\,\sqrt{2}\frac{Y(Y+2) + X}{1 + 2Y^2 +4X}\;,
\label{T20}
\end{equation}
where
$$
Y = \frac{2}{3}\eta\,\frac{G_Q(Q^2)}{G_C(Q^2)}\;,\quad X =
\frac{1}{6}\frac{G_M^2(Q^2)}{G_C^2(Q^2)}\,f(\theta)\;,
$$
$$
f(\theta_e) = 1 + 2(1 + \eta)\,\tan^2\frac{\theta}{2}\;,\quad
\eta = \frac{Q^2}{4M^2_d}\;,
$$
$\theta$ is the scattering angle in the laboratory frame.

In the range of existing experiments one can neglect
$X$ so, that
Eq.(\ref{T20}) takes the form:
\begin{equation}
T_{20}(Q^2) = -\,\sqrt{2}\frac{Y(Y+2)}{1 + 2Y^2}\;.
\label{T20appr}
\end{equation}

To elucidate the role of relativistic effects in
$T_{20}(Q^2)$ let us calculate the quantity
\begin{equation}
\Delta(Q^2) = {T^R_{20}(Q^2)} - {T^{NR}_{20}(Q^2)}\;. \label{DT20}
\end{equation}
Here $T^R_{20}(Q^2)$ is the relativistic value of
$T_{20}(Q^2)$, calculated according to (\ref{T20}),
(\ref{GqGRIP}), and $T^{NR}_{20}(Q^2)$ -- is the
corresponding nonrelativistic value given by (\ref{T20}), (\ref{GqGNIP}).

\begin{figure}
\includegraphics[width=\columnwidth]{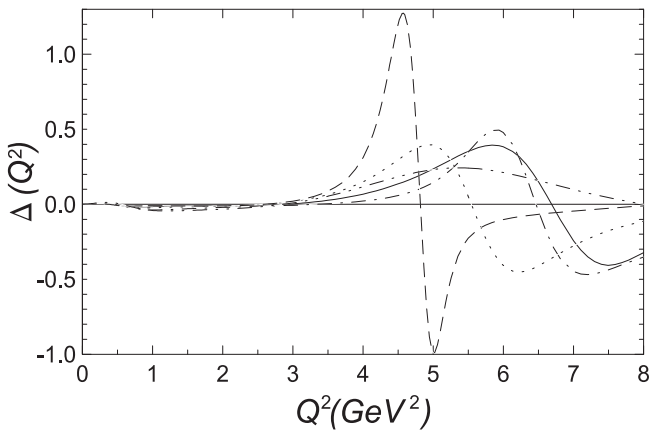}
\caption{
$\Delta(Q^2)$ calculated following Eq.(\protect\ref{Fig.1}) with nucleon
form factors~\protect\cite{GaK71} and different wave functions.
Solid line - N-II~\protect\cite{StK94}, dashed
--~\protect\cite{MuT81}, dotted --~\protect\cite{LaL81}, dot-dashed --
N-I~\protect\cite{GiG02}, dashed double--dotted line
--~\protect\cite{Mac01}.}
\label{Fig.1}
\end{figure}

The dependence of relativistic effects on the choice of the interaction
model is shown in Fig.~\ref{Fig.1}.

The calculation was made using nucleon form factors~\cite{GaK71}
and different model wave functions. One can see from
Fig.~\ref{Fig.1} that the relativistic effects are small for
$Q^2\;\simeq$ 3 GeV$^2$ for all of wave functions.
So, in the region available for the to-day experiment for
$T_{20}(Q^2)$ the relativistic corrections calculated in our approach are
small and almost independent of model wave functions. At  $Q^2\;\ge$ 3.5
GeV$^2$ the corrections become larger and depending upon the model.

Let us discuss the role of the nucleon structure. To estimate this role we
have calculated $T_{20}(Q^2)$ for different fits for nucleon form factors.
Let us note that one of the fits is that given in~\cite{GiG02}
and taking into account recent data for the
ratio of the charge to magnetic form factors for proton
$G^p_E/G^p_M$, obtained in  JLab  experiment
(see, e.g.,~\cite{Jon00}).

\begin{figure}
\includegraphics[width=\columnwidth]{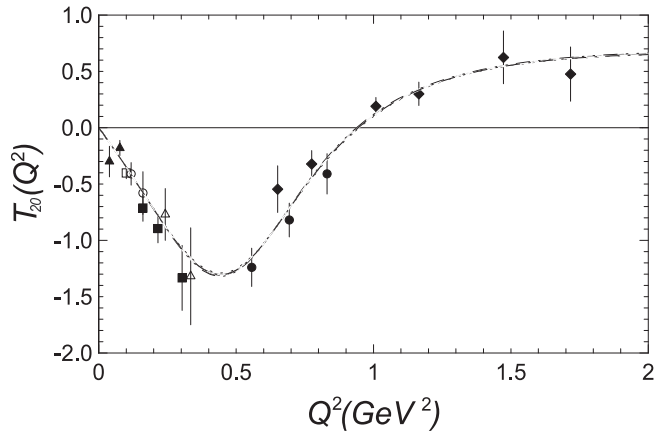}
\caption{$T_{20}(Q^2)$ calculated
with MT wave functions~\protect\cite{MuT81}
and different nucleon form factors.
Solid line -~\protect\cite{GaK85},
dashed --~\protect\cite{GaK71}, dotted --~\protect\cite{MeM96},
dot--dashed --~\protect\cite{GiG02}, dashed
double--dotted --~\protect\cite{Lom01}. The lines are almost
 indistinguishible. Experimental data: open circles
--~\protect\cite{ScB84}, open squares --~\protect\cite{FeB96}, open
triangles --~\protect\cite{GiH90}, filled circles --~\protect\cite{ThA91},
filled squares --~\protect\cite{BoA99}, filled diamonds
--~\protect\cite{AbA00}, filled triangles --~\protect\cite{Dmi85}. }
\label{Fig.2}
\end{figure}

Relativistic $T_{20}(Q^2)$ calculated with the use of different nucleon
form factors and with MT wave functions obtained through the
potentialless approach to inverse scattering problem in~\cite{MuT81}
(see also~\cite{Tro94}) is shown in Fig.~\ref{Fig.2}.

From Fig.\ref{Fig.2}, one can see, as one would expect (see, e.g.,
the discussion in Ref.~\cite{GiG02}), that the dependence on the fit for
nucleon form factors is weak. Note, that this  result does not depend on
the form of wave functions used in the calculation. So,
$T_{20}(Q^2)$ depends weakly on the nucleon structure.

Let us discuss possible contributions to
$T_{20}(Q^2)$ of two--particle MEC.

It is accepted generally that one has to take MEC into account in
a way compatible with the basic principles of the chosen
approach. So, the value of MEC corrections is different for
different approaches. We hope that we can neglect MEC in our
approach when the relativistic corrections are small. The base
for this is given by the following theorem (Siegert,~\cite{Sie37}; see
especially the case of deuteron in~\cite{Bai79}).  If the
electromagnetic current satisfies the conservation law in the
differential form and if the dynamics of the two--particle system
is of nonrelativistic type then the charge density of the exchange current
(the null component) is zero independently of the kind of the potential.
So, in the range of the energy where the nonrelativistic dynamics is valid
(the continuity equation is valid everywhere) the exchange current
contributions to the charge and quadrupole form factors are zero. We
suppose that when the nonrelativistic dynamics is valid approximately then
the MEC contributions to $T_{20}(Q^2)$ are small.

In the experimental range of $Q^2$ the approximate equation (\ref{T20appr})
is valid for
$T_{20}(Q^2)$, so that this quantity is a function of charge and
quadrupole form factors only. This means that MEC contribution to
$T_{20}(Q^2)$ is small.

So, in our approach, the quantity
$T_{20}(Q^2)$ depends weakly on relativistic effects, on meson exchange
currents and on nucleon internal structure. This quantity is defined mainly
by the choice of the deuteron wave function, so that polarization
experiments really could be the test experiments for these wave functions.
One can use the experimental data for $T_{20}(Q^2)$ to choose the most
adequate deuteron wave functions. In fact, we have made calculations using
different model wave functions to compare the predictions with the
experiment. Fig.3 presents the results of our calculation of $T_{20}(Q^2)$
with the use of the different wave
functions~\cite{LaL81,StK94,Mac01,MuT81} and nucleon form factors
from~\cite{GiG02} as well as the experimental points from the
papers~\cite{ScB84,Dmi85,GiH90,ThA91,FeB96,BoA99,AbA00}.

\begin{figure}
\includegraphics[width=\columnwidth]{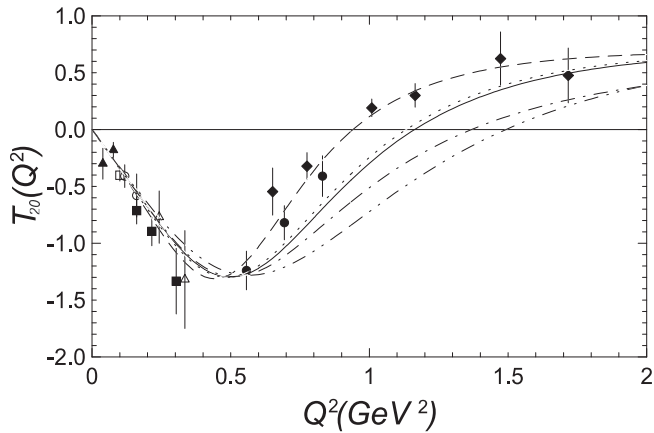}
\caption{Experimental data (legend as in
 Fig.\protect\ref{Fig.1}) and $T_{20}(Q^2)$ calculated with nucleon form
factor~\protect\cite{GaK71} and different wave functions (legend as in
Fig.~\protect\ref{Fig.2}). }
\label{Fig.3}
\end{figure}
The calculation was made with the use of nucleon form factors obtained in
the paper~\cite{GaK71}.
One can see that the results are strongly model dependent and the best
description of experimental data is obtained with the wave functions
obtained by the potentialless approach to inverse scattering
problem~\cite{MuT81}. The results
for different models coincide only at
$Q^2\;\le$ 0.5 GeV$^2$. The recent experimental data~\cite{AbA00}
unambiguously choose the MT wave functions~\cite{MuT81}
in comparison with the model wave functions~\cite{LaL81,StK94,Mac01},
which  are the most largely used now in
nuclear calculations.

The important feature of MT-wave functions is the fact that they
are ``almost model independent'':  no form of $NN$--interaction
Hamiltonian is used. However, the MT wave functions are given by the
dispersion type integral directly in terms of the experimental
scattering phases and the mixing parameter for $NN$--scattering in
the $^3S_1-^3D_1$ channel. Regge--analysis of experimental data
on $NN$--scattering was used to describe the phase shifts at large
energy.

It is worth to notice that the MT wave functions were obtained
using quite general assumptions about analytical properties of
quantum amplitudes such as the validity of Mandelstam
representation for deuteron electrodisintegration amplitude.
These wave functions have no fitting parameters and can be
altered only with the improvement of the $NN$--scattering phase
analysis. The MT wave functions were used in nonrelativistic
calculation of deuteron form factors~\cite{BeM83} and for the
relativistic deuteron structure in~\cite{Arn87}.

Let us notice that the construction of these wave
functions is closely related to the equations obtained in the
framework of the dispersion approach based on the analytic
properties of the scattering
amplitudes~\cite{TrS69,KiT75,AnK92,AnM95}
(see also~\cite{KrT02} and
especially the detailed version~\cite{KrT01h}). In fact, this
approach is a kind of dispersion technique using integrals over
composite--system masses.
Let us note that the MT wave functions have been obtained long in advance
for the polarization experiments  and contain no parameters to
be fitted from deuteron properties.

So, in our approach  the problem of determination
of the behavior of deuteron wave functions at small distances from
polarization experiments is solved. Let us note, that in other approaches
with different dynamics the good description of
$T_{20}(Q^2)$ also can be achieved. However, in those approaches
it seems to be impossible to separate the contributions to $T_{20}(Q^2)$
of the dynamics itself, of relativistic effects generated by the current
operator construction, and effects of nuclear structure. This concerns,
for example, the light--front RHD calculations~\cite{CaK99}.
In the approach ~\cite{CaK99} quite different dynamics is
used which gives 16--component deuteron wave function and good
description of
$T_{20}(Q^2)$ is achieved because of relativistic
corrections.

\section{Conclusion}

In this paper
the deuteron tensor polarization ${T_{20}(Q^2)}$ is calculated
through relativistic Hamiltonian dynamics approach. It is shown  that the
experimental data for $T_{20}(Q^2)$ component of deuteron polarization
tensor in elastic electron--deuteron scattering up to $Q^2 \approx
2(GeV/c)^2$ can be described in terms of nonrelativistic theory with no
account  of relativistic effects and meson exchange currents. These data
for $T_{20}(Q^2)$ could be a touchstone for nonrelativistic deuteron wave
functions, the results of calculations depending crucially on the choice of
wave functions. It is also shown that the wave functions obtained by the
dispersion method of potentialless inverse scattering problem give the
best results for $T_{20}(Q^2)$.

\begin{widetext}
\appendix{}
\section{Relativistic free two--nucleon form factors in
$^3S_1-^3D_1$ - channel}
Relativistic two--particle form factors of free (without interaction)
$np$- system in the $^3S_1-^3D_1$--channel are $2\times2$ matrices. The
elements of three corresponding matrices are given below.

Free two-particle charge form--factor (see below for the notations):
$$
g^{ll'}_{0C}(s, Q^2, s') = R(s, Q^2, s')\,Q^2 \left[\,(s + s' +
Q^2)\left(G^p_E(Q^2)+G^{n} _E(Q^2)\right)g^{ll'}_{CE} + \right.
$$
$$
\left. + \frac{1}{M}\xi(s,Q^2,s')
\left(G^p_M(Q^2)+G^{n}_M(Q^2)\right)g^{ll'}_{CM} \right]\;,
$$
$$
g^{00}_{CE} = \left(\frac{1}{2}\cos\omega_1\cos\omega_2 +
\frac{1}{6}\sin\omega_1\sin\omega_2\right)\;,\quad g^{00}_{CM} =
\left(\frac{1}{2}\cos\omega_1\sin\omega_2 -
\frac{1}{6}\sin\omega_1\cos\omega_2\right)\;,
$$
$$
g^{02}_{CE} = -\frac{1}{6\sqrt{2}}\, \left(P'_{22} +
2\,P'_{20}\right)\sin\omega_1\sin\omega_2\;,\quad g^{02}_{CM} =
\frac{1}{6\sqrt{2}}\, \left(P'_{22} +
2\,P'_{20}\right)\sin\omega_1\cos\omega_2\;,
$$
$$
g^{22}_{CE} = \left[\frac{1}{2}\,L_1\,\cos\omega_1\cos\omega_2 +
\frac{1}{24}\,L_2\,\sin(\omega_2 - \omega_1) +
\frac{1}{12}\,L_3\,\sin\omega_1\sin\omega_2\right]\;,
$$
\begin{equation}
g^{22}_{CM} =
-\,\left[-\frac{1}{2}\,L_1\,\cos\omega_1\sin\omega_2 +
\frac{1}{24}\,L_2\,\cos(\omega_2 - \omega_1) +
\frac{1}{12}\,L_3\,\sin\omega_1\cos\omega_2\right]\;. \label{g0C}
\end{equation}

Quadrupole two-particle charge form--factor:
$$
g^{ll'}_{0Q}(s, Q^2, s') = \frac{1}{2}R(s, Q^2, s')\,Q^2
\left[\,(s + s' + Q^2)\left(G^p_E(Q^2)+G^{n}
_E(Q^2)\right)g^{ll'}_{QE} + \right.
$$
$$
\left. + \frac{1}{M}\xi(s,Q^2,s')
\left(G^p_M(Q^2)+G^{n}_M(Q^2)\right)g^{ll'}_{QM} \right]\;,
$$
$$
g^{00}_{QE} = \sin\omega_1\sin\omega_2\;,\quad g^{00}_{QM} =
-\,\sin\omega_1\cos\omega_2\;,
$$
$$
g^{02}_{QE} = -\,\frac{3}{2\sqrt{2}}\left\{
2\,P'_{20}\cos\omega_1\cos\omega_2 - P'_{21}\sin(\omega_1 -
\omega_2) + \frac{1}{3}\left(4\,P'_{20} - P'_{22}\right)
\sin\omega_1\sin\omega_2\right\}\;,
$$
$$
g^{02}_{QM} = \frac{3}{2\sqrt{2}}\left\{
-\,2\,P'_{20}\cos\omega_1\sin\omega_2 + P'_{21}\cos(\omega_1 -
\omega_2) + \frac{1}{3}\left(4\,P'_{20} - P'_{22}\right)
\sin\omega_1\cos\omega_2\right\}\;,
$$
$$
g^{22}_{QE} = \frac{3}{2}\left\{ \,L_4\,\cos\omega_1\cos\omega_2
-\,\frac{1}{12}\,L_5\,\sin(\omega_1 - \omega_2) +
\frac{1}{6}\,L_6\,\sin\omega_1\sin\omega_2\right\}\;,
$$
\begin{equation}
g^{22}_{QM} = -\,\frac{3}{2}\,\left\{
-\,L_4\,\cos\omega_1\sin\omega_2 +
\frac{1}{12}\,L_5\,\cos(\omega_1 - \omega_2) +
\frac{1}{6}\,L_6\,\sin\omega_1\cos\omega_2\right\}\;, \label{g0Q}
\end{equation}

Magnetic two-particle charge form--factor:
$$
g^{ll'}_{0M}(s, Q^2, s') = -\,R(s, Q^2, s')\,
\left[\,\xi(s,Q^2,s')\left(G^p_E(Q^2)+G^{n}
_E(Q^2)\right)g^{ll'}_{ME} + \right.
$$
$$
\left. + \left(G^p_M(Q^2)+G^{n}_M(Q^2)\right)g^{ll'}_{MM}
\right]\;,
$$
$$
g^{00}_{ME} = \sin(\omega_1-\omega_2)\;,
$$
$$
g^{00}_{MM} = \frac{1}{2\,M}\left\{ \left[\gamma_1 -
\frac{1}{2}\left(\gamma_3(s,Q^2,s') +
\gamma_3(s',Q^2,s)\right)\right]\cos\omega_1\cos\omega_2 + \right.
$$
$$
\left. + \frac{1}{4}\left(\gamma_2(s,Q^2,s') +
\gamma_2(s',Q^2,s)\right) \cos\omega_1\sin\omega_2 +
\frac{1}{2}\gamma_1\,\sin\omega_1\sin\omega_2\right\}\;,
$$
$$
g^{02}_{ME} = -\,\frac{1}{4\sqrt{2}} \left(P'_{22} +
2\,P'_{20}\right)\sin(\omega_1-\omega_2)\;,
$$
$$
g^{02}_{MM} = \frac{1}{8\sqrt{2}\,M}\left\{
-\,\left[2\,P'_{20}\gamma_1 + P'_{21}\gamma_2 + \left(P'_{22} -
2\,P'_{20}\right)\gamma_3\right] \cos\omega_1\cos\omega_2 +
\right.
$$
$$
+ \left[P'_{21}\gamma_1 + \frac{1}{2}\left(P'_{22} -
2\,P'_{20}\right)\gamma_2 - 2\,P'_{21}\gamma_3\right]
\cos\omega_1\sin\omega_2 +
$$
$$
+ \left[P'_{21}\gamma_1 - \frac{1}{2}\left(P'_{22} +
6\,P'_{20}\right)\gamma_2 - 2\,P'_{21}\gamma_3\right]
\sin\omega_1\cos\omega_2 +
$$
$$
\left. + \left[2\,P'_{20}\gamma_1 + P'_{21}\gamma_2 -
\left(P'_{22} + 6\,P'_{20}\right)\gamma_3\right]
\sin\omega_1\sin\omega_2\right\}\;,
$$
$$
g^{22}_{ME} = -\,\frac{1}{4}\left\{ \frac{1}{2}L_2\,\cos(\omega_1
- \omega_2) + L_3\,\sin(\omega_1-\omega_2)\right\}\;,
$$
$$
g^{22}_{MM} = \frac{1}{8\,M}\left\{ \left[-\,L_7\,\gamma_1 -
\frac{1}{8}L_8\,\left(\gamma_2(s,Q^2,s') -
\gamma_2(s',Q^2,s)\right) + \right.\right.
$$
$$
\left. + \frac{1}{2}L_9\, \left(\gamma_3(s,Q^2,s) +
\gamma_3(s',Q^2,s)\right)\right] \cos\omega_1\cos\omega_2 +
$$
$$
+ \frac{1}{4}\left[\left(L_{10}(s,Q^2,s') +
L_{10}(s',Q^2,s)\right) \gamma_1 -\,L_9\,
\left(\gamma_2(s,Q^2,s') + \gamma_2(s',Q^2,s)\right) -\right.
$$
$$
\left. -\,L_8\,\left(\gamma_3(s,Q^2,s') -
\gamma_3(s',Q^2,s)\right)\right] \cos\omega_1\sin\omega_2 +
$$
$$
+ \frac{1}{4}\left[8\,L_{11}\,\gamma_1 + L_{12}\,
\left(\gamma_2(s,Q^2,s') - \gamma_2(s',Q^2,s)\right) + \right.
$$
$$
\left. + L_{13}\, \left(\gamma_3(s,Q^2,s') +
\gamma_3(s',Q^2,s)\right)\right] \sin\omega_1\cos\omega_2 +
$$
$$
+ \frac{1}{2}\left[ \left(L_{14}(s,Q^2,s') +
L_{14}(s',Q^2,s)\right)\gamma_1 -\,\frac{1}{4}L_{13}\,
\left(\gamma_2(s,Q^2,s') + \gamma_2(s',Q^2,s)\right) + \right.
$$
\begin{equation}
\left.\left. + L_{12}\, \left(\gamma_3(s,Q^2,s') -
\gamma_3(s',Q^2,s)\right)\right]
\sin\omega_1\sin\omega_2\right\}\;. \label{g0M}
\end{equation}

The following equation is valid for form factors:
$$
g^{ll'}_{0i}(s, Q^2, s') = g^{l'l}_{0i}(s', Q^2, s)\;,\quad
i=C,Q,M\;.
$$

Notations:
$$
R(s, Q^2, s') = \frac{(s+s'+Q^2)}{\sqrt{(s-4M^2) (s'-4M^2)}}\,
\frac{\vartheta(s,Q^2,s')}{{[\lambda(s,-Q^2,s')]}^{3/2}}
\frac{1}{\sqrt{1+Q^2/4M^2}}\;,
$$
$$
\xi(s,Q^2,s')=\sqrt{ss'Q^2-M^2\lambda(s,-Q^2,s')}\>,
$$
$$
L_1 = L_1(s,Q^2,s') = P_{20}\,P'_{20} +
\frac{1}{3}P_{21}\,P'_{21} + \frac{1}{12}P_{22}\,P'_{22}\;,
$$
$$
L_2 = L_2(s,Q^2,s') = P_{21}\left(P'_{22} - 6\,P'_{20}\right) -
P'_{21}\left(P_{22} - 6\,P_{20}\right)\;,
$$
$$
L_3 = L_3(s,Q^2,s') = 2\,P_{21}\,P'_{21} + 4\,P_{20}\,P'_{20} -
P_{20}\,P'_{22} - P_{22}\,P'_{20}\;,
$$
$$
L_4 = L_4(s,Q^2,s') = P_{20}\,P'_{20} +
\frac{1}{6}\,P_{21}\,P'_{21} - \frac{1}{12}P_{22}\,P'_{22}\;,
$$
$$
L_5 = L_5(s,Q^2,s') = P'_{21}\left(P_{22} + 6\,P_{20}\right) -
P_{21}\left(P'_{22} + 6\,P'_{20}\right)\;,
$$
$$
L_6 = L_6(s,Q^2,s') = 8\,P_{20}\,P'_{20} + P_{21}\,P'_{21} +
P_{20}\,P'_{22} + P_{22}\,P'_{20}\;,
$$
$$
L_7 = L_7(s,Q^2,s') = P_{21}\,P'_{21} + 4\,P_{20}\,P'_{20}\;,
$$
$$
L_8 = L_8(s,Q^2,s') = P_{21}\left(P'_{22} + 2\,P'_{20}\right) +
P'_{21}\left(P_{22} +2\,P_{20}\right)\;,
$$
$$
L_9 = L_9(s,Q^2,s') = P_{20}\,P'_{22} + P_{22}\,P'_{20} +
4\,P_{20}\,P'_{20}\;,
$$
$$
L_{10} = L_{10}(s,Q^2,s') = P_{22}\,P'_{21} + 4\,P_{21}\,P'_{20} -
2\,P_{20}\,P'_{21}\;,
$$
$$
L_{11} = L_{11}(s,Q^2,s') = P'_{21}\,P_{20} - P'_{20}\,P_{21}\;,
$$
$$
L_{12} = L_{12}(s,Q^2,s') = P_{20}\,P'_{22} - P_{22}\,P'_{20}\;,
$$
$$
L_{13} = L_{13}(s,Q^2,s') = P_{21}\left(P'_{22} +
2\,P'_{20}\right) - P'_{21}\left(P_{22} +2\, P_{20}\right)\;,
$$
$$
L_{14} = L_{14}(s,Q^2,s') = P_{22}\,P'_{20} - P_{21}\,P'_{21} -
2\,P_{20}\,P'_{20}\;,
$$
$$
\gamma_1 = \gamma_1(s,Q^2,s') = (s + Q^2 + s')\,Q^2\;,
$$
$$
\gamma_2 = \gamma_2(s,Q^2,s') = \xi(s,Q^2,s') \frac{(s - s'
+Q^2)(\sqrt{s'} + 2\,M) + (s' - s + Q^2)\sqrt{s'}}
{\sqrt{s'}(\sqrt{s'} + 2\,M)}\;,
$$
$$
\gamma_3 = \gamma_3(s,Q^2,s') =
\frac{\xi^2(s,Q^2,s')}{\sqrt{s'}(\sqrt{s'} + 2\,M)}\;.
$$
$\omega_1$ and $\omega_2$ -- the Wigner rotation parameters,
$$
\omega_1 =
\arctan\frac{\xi(s,Q^2,s')}{M\left[(\sqrt{s}+\sqrt{s'})^2 +
Q^2\right] + \sqrt{ss'}(\sqrt{s} +\sqrt{s'})}\>,
$$
\begin{equation}
\omega_2 = \arctan\frac{ \alpha (s,s') \xi(s,Q^2,s')} {M(s + s' +
Q^2) \alpha (s,s') + \sqrt{ss'}(4M^2 + Q^2)}\>, \label{omega}
\end{equation}
where $\alpha (s,s') = 2M + \sqrt{s} + \sqrt{s'}$. $P_{2i} =
P_{2i}(z)\;,\;P'_{2i} = P_{2i}(z')$ -- adjoint Legendre functions:
\begin{equation}
P_{20}(z) = \frac{1}{2}\left(3\,z^2 -1\right)\;,\quad P_{21}(z) =
3\,z\sqrt{1 - z^2}\;,\quad P_{22}(z) = 3\left(1 - z^2\right)\;.
\label{Leg}
\end{equation}
$z\;,\;z'$ -- the arguments of Legendre functions:
$$
z = z(s,Q^2,s') = \frac{\sqrt{s}(s' - s -
Q^2)}{\sqrt{\lambda(s,-Q^2,s')(s - 4\,M^2)}}\;,\quad z' =
z'(s,Q^2,s') = -\,z(s',Q^2,s)\;.
$$
$\vartheta(s,Q^2,s')= \theta(s'-s_1)-\theta(s'-s_2)$, $\theta$ -
step function.
$$
s_{1,2}=2M^2+\frac{1}{2M^2} (2M^2+Q^2)(s-2M^2) \mp \frac{1}{2M^2}
\sqrt{Q^2(Q^2+4M^2)s(s-4M^2)}\;.
$$
The functions $s_{1,2}(s,Q^2)$ give the kinematically available region
in the plane $(s,s')$. They are obtained in~\cite{KrT02}.
$G^{p,n}_{E,M}(Q^2)$ are Sachs form
factors of proton and neutron.

\section{Nonrelativistic free two--nucleon form factors in
$^3S_1-^3D_1$ - channel}

Nonrelativistic charge two--particle free form factor:
$$
\tilde g^{ll'}_{0C}(k, Q^2, k') = g(k, Q^2, k')
\left(G^p_E(Q^2)+G^{n}_E(Q^2)\right)\tilde g^{ll'}_{CE} \;,
$$
$$
\tilde g^{00}_{CE} = 1\;,\quad \tilde g^{02}_{CE} = \tilde
g^{20}_{CE} = 0\;,
$$
\begin{equation}
\tilde g^{22}_{CE} = \tilde P_{20}\,\tilde P'_{20} +
\frac{1}{3}\tilde P_{21}\,\tilde P'_{21} + \frac{1}{12}\tilde
P_{22}\,\tilde P'_{22}\;, \label{g0Cn}
\end{equation}
Nonrelativistic quadrupole two--particle free form factor:
$$
\tilde g^{ll'}_{0Q}(k, Q^2, k') = \frac{3}{2}g(k, Q^2, k')
\left(G^p_E(Q^2)+G^{n}_E(Q^2)\right)\tilde g^{ll'}_{QE} \;,
$$
$$
\tilde g^{00}_{QE} = 0\;,\quad \tilde g^{02}_{QE} =
-\,\sqrt{2}\,\tilde P'_{20}\;,\quad \tilde g^{20}_{QE} =
-\,\sqrt{2}\,\tilde P_{20}\;,
$$
\begin{equation}
\tilde g^{22}_{QE} = \tilde P_{20}\,\tilde P'_{20} +
\frac{1}{6}\tilde P_{21}\,\tilde P'_{21} - \frac{1}{12}\tilde
P_{22}\,\tilde P'_{22}\;, \label{g0Qn}
\end{equation}

Nonrelativistic magnetic two--particle free form factor:
$$
\tilde g^{ll'}_{0M}(k, Q^2, k') = -\,\frac{1}{4\,\sqrt{2}\,M}g(k,
Q^2, k') \left[\left(G^p_E(Q^2)+G^{n}_E(Q^2)\right)\tilde
g^{ll'}_{ME} + \right.
$$
$$
\left. + \left(G^p_M(Q^2)+G^{n}_M(Q^2)\right)\tilde
g^{ll'}_{MM}\right].
$$

$$
\tilde g^{00}_{ME} = \tilde g^{02}_{ME} = \tilde g^{20}_{ME} =
0\;,
$$
$$
\tilde g^{00}_{MM} = 4\,\sqrt{2}\;,\quad g^{02}_{MM} =
-\,{2}\,\tilde P'_{20}\;,\quad g^{20}_{MM} = -\,{2}\,\tilde
P_{20}\;,
$$
$$
\tilde g^{22}_{ME} =\gamma\left[\tilde P_{21}\, \left(\tilde
P'_{22} - 6\,\tilde P'_{20}\right) - \tilde P'_{21}\,
\left(\tilde P_{22} - 6\,\tilde P_{20}\right)\right]\,
$$
\begin{equation}
g^{22}_{MM} =-\,\sqrt{2}\left[\tilde P_{21}\,\tilde P'_{21} +
4\,\tilde P'_{20}\,\tilde P_{20}\right]\;. \label{g0Mn}
\end{equation}
Here
$$
g(k,Q^2,k') = \frac{1}{k\,k'\,Q} \left[\theta\left(k' - \left|k -
\frac{Q}{2}\right|\right) - \theta\left(k' - k -
\frac{Q}{2}\right)\right]\;.
$$
$\tilde P_{2i} = P_{2i}(y)\;,\;\tilde P'_{2i} = P_{2i}(y')$ --
adjoint Legendre functions
(\ref{Leg}). $y\;,\;y'$ --
the arguments of Legendre functions:
$$
y = y(k,Q^2,k') = -\,\frac{4(k^2 - k'\,^2) +Q^2}{4\,k\,Q}\;,\quad
y' = y'(k,Q^2,k') = -\,y(k',Q^2,k)\;.
$$
$$
\gamma = -\,\frac{1}{2} \left(\frac{2k'\,^2(1 -
y'\,^2)}{Q^2}\right)^{1/2} = -\,\frac{1}{2} \left(\frac{2k^2(1 -
y^2)}{Q^2}\right)^{1/2} \;.
$$

              \end{widetext}
\end{document}